\documentclass[osajnl,twocolumn,showpacs,superscriptaddress,10pt]{revtex4-1} 

\pdfoutput=1

\usepackage{amsmath,amssymb,graphicx}
\usepackage[colorlinks=true, allcolors=blue]{hyperref}

\begin{document}

\title{Squeezing-enhanced measurement sensitivity in a cavity optomechanical system}

\author{Hugo Kerdoncuff}\email{Corresponding author: hk@dfm.dk}
\affiliation{Department of Physics, Technical University of Denmark, Fysikvej bld. 309, Lyngby 2800 Kgs., Denmark}

\author{Ulrich B. Hoff}
\affiliation{Department of Physics, Technical University of Denmark, Fysikvej bld. 309, Lyngby 2800 Kgs., Denmark}

\author{Glen I.Harris}
\affiliation{Centre of Excellence in Engineered Quantum Systems, University of Queensland, St Lucia, Queensland 4072, Australia}

\author{Warwick P. Bowen}
\affiliation{Centre of Excellence in Engineered Quantum Systems, University of Queensland, St Lucia, Queensland 4072, Australia}

\author{Ulrik L. Andersen}
\affiliation{Department of Physics, Technical University of Denmark, Fysikvej bld. 309, Lyngby 2800 Kgs., Denmark}



\begin{abstract}
We determine the theoretical limits to squeezing-enhanced measurement sensitivity of mechanical motion in a cavity optomechanical system. The motion of a mechanical resonator is transduced onto quadrature fluctuations of a cavity optical field and a measurement is performed on the optical field exiting the cavity. We compare measurement sensitivities obtained with coherent probing and quantum-enhanced probing of the mechanical motion, i.e. the coherent probe field carries vacuum states and quadrature squeezed vacuum states at sideband frequencies, respectively. We find that quantum-enhanced probing provides little to no improvement in motion sensing for resonators in the unresolved sideband regime but may significantly increase measurement sensitivities for resonators in the resolved sideband regime.

\end{abstract}


\maketitle

\section{Introduction}

High-sensitivity measurements of mechanical displacements were initially conducted in large scale interferometers for the detection of gravitational waves predicted by Einstein. From then on, coupling between mechanical oscillators and optical fields via radiation pressure has been extensively studied within the field of optomechanics and has found applications for single spin detection by magnetic resonance force microscopy \cite{Rugar2004}, attometer-scale displacement measurements \cite{Arcizet2006, Schliesser2008}, chip-based room temperature magnetometry \cite{Forstner2012, Forstner2014}, and observation of quantum phenomena in macroscopic mechanical systems \cite{Teufel2011, Safavi2012, Verhagen2012, Palomaki2013}. Continuous measurement of the displacement of a mechanical resonator involves the transduction of mechanical motion onto the amplitude and phase fluctuations of an optical probe beam which are subsequently measured with an interferometer. Such a measurement is limited by imprecision noise set by the photon shot noise of the probe light and scaling as $1/\sqrt{N}$ where N is the number of photons in the probe light
. It seems that arbitrarily precise readout can be achieved simply by increasing the optical power, however a fluctuating radiation pressure force imparted on the mechanical resonator due to photon shot noise gives rise to quantum back-action noise which scales as $\sqrt{N}$ and counteracts the improvement in measurement sensitivity for large $N$. The limit in readout precision achieved when the sum of the two noise contributions is minimum, is called the standard quantum limit (SQL) \cite{Caves1980, Braginsky1992}. Imprecision noise surpassing the SQL has recently been achieved in a regime where the total measurement noise is dominated by mechanical thermal fluctuations \cite{Anetsberger2009, Teufel2009, Westphal2012}. This was quickly followed by experiments where radiation pressure noise dominates the thermal noise, that allowed the observation of radiation-pressure shot noise and ponderomotive squeezing \cite{Purdy2013, Safavi2013}. Furthermore, experiments with atom clouds cooled close to their motional ground state have recently demonstrated force sensing within a factor of 4 above the SQL \cite{Schreppler2014}.

Since the early days of large-scale gravitational-wave interferometers, it was recognized that the use of squeezed light as a quantum sensing resource could improve the sensitivity of interferometric phase measurements beyond the shot noise limit \cite{Caves1981}. This was later experimentally demonstrated for Mach-Zehnder \cite{Min1987}, Michelson \cite{McKenzie2002, Goda2008}, Sagnac \cite{Eberle2010} and large-scale gravitational-wave interferometers \cite{Ligo2011}, and applied to displacement measurements \cite{Treps2002} of particles in biological samples \cite{Taylor2013} and micro-mechanical resonators \cite{Hoff2013}.
The use of squeezed light 
gives access to SQL sensitivity at lower optical probe power, as a reduction of imprecision noise on the squeezed quadrature is accompanied by an increase in quantum back-action noise due to antisqueezing on the conjugate quadrature. The use of squeezed light is therefore beneficial in systems for which the use of high optical powers is limited by technical reasons, e.g. maximum output power of existing lasers, or practical reasons, e.g. damage threshold of biological samples. In high-Q microcavity resonators, increasing optical powers gives rise to additional noise due to parametric instabilities \cite{Rokhsari2005} what prevents reaching sensitivities at the SQL. The aim of this article is to determine the conditions for reaching optimum quantum-enhanced displacement sensitivities in cavity optomechanical systems.

\section{Optomechanical equations of motion}

Cavity optomechanical systems \cite{Aspelmeyer2013} come in a lot of different arrangements, from cold atoms trapped in optical cavities \cite{Murch2008, Brennecke2008} to nano- \cite{Eichenfield2009} and micro-scale \cite{Thompson2008, Anetsberger2008, Wilson2009} resonators, and gram-scale mirrors in interferometers \cite{Corbitt2007}. They consist of an optical cavity mode coupled to a mechanical harmonic oscillator via the radiation pressure mediated optomechanical interaction. The motion of the mechanical oscillator modifies the optical length of the cavity resulting in a shift of the optical resonance frequency. This shift is typically assumed to be linear under small mechanical displacements. Reciprocally, radiation pressure forces from the light circulating in the cavity affects the position of the mechanical oscillator.
Formally, the optomechanical system is described by the following Hamiltonian,
\begin{equation}\label{system_hamiltonian}
\hat{H}=\hbar\omega_{0}\hat{a}^{\dagger}\hat{a}+\hbar\Omega_{m}\hat{b}^{\dagger}\hat{b}+\hbar g_{0}\hat{x}\hat{a}^{\dagger}\hat{a}+\hat{H}_{\kappa}+\hat{H}_{\Gamma},
\end{equation}
where the cavity (oscillator) field is represented by annihilation and creation operators $\hat{a}$ ($\hat{b}$) and $\hat{a}^{\dagger}$ ($\hat{b}^{\dagger}$), respectively, and resonates at frequency $\omega_0$ ($\Omega_m$). The first (second) term in Eq. \ref{system_hamiltonian} denotes the energy stored in the optical cavity (mechanical oscillator), and the fourth (fifth) term comprises the exchange of energy between the cavity (oscillator) field and external photon (phonon) baths. The third term represents the parametric coupling of the oscillator's position, $\hat{x}=\hat{b}+\hat{b}^{\dagger}$, with the intracavity field. The optomechanical interaction scales with the vacuum optomechanical coupling parameter $g_0$ \cite{Gorodetsky2010}
, which quantifies the optical frequency noise arising from the zero-point-motion of the oscillator.

The optomechanical interaction involves the nonlinear mixing of optical and mechanical field operators and may be linearized by considering that the cavity is driven by a dominant driving field at frequency $\omega_d$ which steers the intracavity field and the mechanical oscillator into a stable equilibrium position $(\bar{a}, \bar{x})$: $\hat{a}\rightarrow\bar{a}+\hat{a}$, $\hat{x}\rightarrow\bar{x}+\hat{x}$. Therefore, the nonlinear mixing of the small optical and mechanical field fluctuations can be neglected and the optomechanical interaction Hamiltonian (third term in Eq. (\ref{system_hamiltonian})) simplifies to:
\begin{equation}\label{int_hamiltonian}
\hat{H}_{int}=\hbar g_{0}\bar{a}\left[\bar{a}\bar{x}+\bar{a}\hat{x}+\bar{x}\hat{X}_{a}+\hat{x}\hat{X}_{a}\right],
\end{equation}
where we have chosen the phase of the intracavity field to be the zero-phase reference, i.e. $\bar{a}$ is real. In Eq. (\ref{int_hamiltonian}) we have introduced the amplitude quadrature operator of the intracavity field fluctuations, $\hat{X}_{a}=\hat{a}+\hat{a}^{\dagger}$, whose conjugate, phase quadrature operator reads, $\hat{P}_{a}=-i\hat{a}+i\hat{a}^{\dagger}$. Similarly, the dimensionless position and momentum operators of the mechanical oscillator read, $\hat{x}=\hat{b}+\hat{b}^{\dagger}$ and $\hat{p}=-i\hat{b}+i\hat{b}^{\dagger}$, respectively. Using the standard Heisenberg picture formalism in a frame corotating with the driving optical field at $\omega_d$  provides the time evolution of the system observables:
\begin{align}
\begin{array}{c} \Dot{\hat{X}}_{a} \\ \Dot{\hat{P}}_{a} \end{array} & \begin{array}{l} = -\frac{\kappa}{2}\hat{X}_{a} - \Delta \hat{P}_{a} + \hat{X}_{in}^{\prime}, \\ = -\frac{\kappa}{2}\hat{P}_{a} + \Delta \hat{X}_{a} + \hat{P}_{in}^{\prime}- g\hat{x} , \end{array} \label{OQLEt}\\
\begin{array}{c} \Dot{\hat{x}} \\ \Dot{\hat{p}} \end{array} & \begin{array}{l} = \Omega_m\hat{p},\\ =  -\Omega_m\hat{x} -\Gamma_m\hat{p} -g\hat{X}_{a}+\hat{F}_{ex}. \end{array} \label{MQLEt}
\end{align}

Optical and mechanical inputs to the optomechanical system are represented by conjugate optical field quadratures $\hat{X}_{in}^{\prime}$ and $\hat{P}_{in}^{\prime}$, and by an external force $\hat{F}_{ex}$, respectively. The prime superscript indicates that the field quadratures are defined in the phase space of the reference field, i.e. the intracavity field. We consider a simple model with two optical inputs, $\hat{X}_{in}^{\prime}=\sqrt{\kappa_{ex}}\hat{X}_{d}^{\prime}+\sqrt{\kappa_{c}}\hat{X}_{vac}^{\prime}$ (and $\hat{P}_{in}^{\prime}=\sqrt{\kappa_{ex}}\hat{P}_{d}^{\prime}+\sqrt{\kappa_{c}}\hat{P}_{vac}^{\prime}$), 
where the two contributions are associated with intracavity loss
by the coupling of the intracavity photon field with an external vacuum field, $\hat{s}_{vac}=\left(\hat{X}_{vac}^{\prime}+i \hat{P}_{vac}^{\prime}\right)/2$, at a rate $\kappa_c$, and with driving field fluctuations, $\hat{s}_{d}=\left(\hat{X}_{d}^{\prime}+i \hat{P}_{d}^{\prime}\right)/2$, exciting the intracavity photon field at a rate $\kappa_{ex}$. The total coupling rate of the optical cavity with its environment is the sum of individual coupling rates to each external photon baths, i.e. $\kappa=\kappa_{c}+\kappa_{ex}$.
The detuning of the driving field from the cavity resonance is  denoted $\Delta=\omega_{d}-\omega_{0}^{\prime}$, where $\omega_{0}^{\prime}$ includes the static frequency shift of the cavity resonance, $g_0 \bar{x}$, induced by the radiation pressure force, $\hbar g_0 \bar{a}^2$, exerted by the driven intracavity optical field. The effective optomechanical coupling rate, $g=2g_{0}\bar{a}$, captures the mutual transduction between the photon and phonon fields : the cavity photons exchange momentum with the mechanical oscillator via the radiation pressure interaction, while the phase of the intracavity field is modulated by the fluctuations of the cavity length. The strength of the optomechanical coupling scales as the square-root of the mean intracavity photon number, $\bar{a}=\sqrt{N_c}$.
In addition to the radiation pressure force, the mechanical oscillator is driven by external forces, $\hat{F}_{ex}$, which may include, for example, thermal Langevin forces from the non-zero environmental temperature, magnetic forces \cite{Rugar2004}, or dielectric gradient forces \cite{Unterreithmeier2009, Lee2010}. The use of the dimensionless mechanical operators $\hat{x}$ and $\hat{p}$ in Eq. (\ref{MQLEt}) results in the radiation pressure force, $\hat{F}_{rad}=-g\hat{X}_{a}$, and the external forces, $\hat{F}_{ex}$, being normalized to the zero-point momentum $p_{zpf}=\hbar/2x_{zpf}$, where $x_{zpf}=\sqrt{\hbar/2\Omega_m M}$ is the zero-point motion of the mechanical oscillator with mass $M$. The zero-point motion and zero-point momentum are the amplitudes of the mechanical oscillator's displacement and momentum when in its ground state.

As the cavity is driven by an harmonic optical field, it is convenient to switch to frequency domain in order to solve equations (\ref{OQLEt}) and (\ref{MQLEt}). By using the following definition for the  Fourier transform of a time-dependent observable, $\tilde{X}[\Omega]=\int^{+\infty}_{-\infty}\hat{X}(t)e^{i\Omega t}dt$, the time-dependent equations (\ref{OQLEt}) and (\ref{MQLEt}) transform into:
\begin{align}
\begin{array}{c} \tilde{X}_{a} \\ \tilde{P}_{a} \end{array} &
\begin{array}{l} = \chi_c \left[\left(\frac{\kappa}{2}-i\Omega\right) \tilde{X}_{in}^{\prime}-\Delta \tilde{P}_{in}^{\prime}+\Delta g\tilde{x}\right],
\\ =\chi_c \left[\Delta \tilde{X}_{in}^{\prime}+\left(\frac{\kappa}{2}-i\Omega\right) \tilde{P}_{in}^{\prime}- \left(\frac{\kappa}{2}-i\Omega\right) g\tilde{x}\right],\end{array}
\label{OQLEf}\\
\begin{array}{c} \tilde{x} \\ \tilde{p} \end{array} &
\begin{array}{l} = \Omega_m \chi_m \left[-g \tilde{X}_{a} + \tilde{F}_{ex}\right],
\\ =-i \Omega\chi_m \left[-g \tilde{X}_{a} + \tilde{F}_{ex}\right],\end{array}\label{MQLEf}
\end{align}
where we introduce an optical and mechanical susceptibility, $\chi_{c}^{-1}=(\kappa/2-i\Omega)^{2}+\Delta^{2}$ and $\chi_m^{-1}=\Omega_{m}^{2}-\Omega^{2}-i\Omega\Gamma_{m}$, respectively.
Equations (\ref{OQLEf}) ((\ref{MQLEf})) shows that the transduced phonon (photon) field adds to contributions from external photon (phonon) baths to drive the intracavity photon (phonon) field before being filtered by the linear response of the cavity (oscillator).

\section{Cavity output}

The dynamics of the photon and phonon field fluctuations of the optomechanical system driven by external photon and phonon field fluctuations is described by the set of equations (\ref{OQLEf}) and (\ref{MQLEf}). In practice however, optomechanical system observables, that are the amplitude and phase quadratures of the intracavity photon field and the position and momentum of the mechanical resonator, are not easily accessible and are inferred via the detection of the cavity output field fluctuations, $\hat{s}_{out}=\left(\hat{X}_{out}^{+}+i\hat{X}_{out}^{-}\right)/2$. Considering a lossy optical cavity driven by an optical field, time reversal symmetry and energy conservation \cite{Haus1984} yields the following linear relations between input, output and intracavity fields:
\noindent\begin{minipage}[b]{0.4\linewidth}
\begin{equation}\bar{a}=\frac{\sqrt{\kappa_{ex}}}{\kappa/2-i\Delta}\bar{s}_d, \label{input-intra}\end{equation}
\end{minipage}%
\begin{minipage}[b]{0.15\linewidth}
~
\end{minipage}%
\begin{minipage}[b]{0.45\linewidth}
\begin{equation}\sqrt{\kappa_{ex}}\hat{a}=\hat{s}_{d} +\epsilon\hat{s}_{out}.\label{input-output}\end{equation}
\end{minipage}\par\vspace{\belowdisplayskip}
\noindent Here, we denote the mean classical parts of the probe and output fields by $\bar{s}_d$ and $\bar{s}_{out}$, respectively.
The coefficient $\epsilon=\pm 1$ can be arbitrarily chosen to provide a physical meaning to Eq. (\ref{input-output}). Common usage sets $\epsilon=1$ for a Fabry-P\'erot-type cavity while $\epsilon=-1$ for a whispering-gallery mode cavity.

As $\bar{a}$ was chosen to carry the zero-phase reference the phases of input and output fields are defined relative to the phase of the intracavity field and can be derived from equations (\ref{input-intra}) and (\ref{input-output}) : 
$\phi_{d}=-\arctan\left[\frac{2\Delta}{\kappa}\right]$, and $\phi_{out}=-\arctan\left[\frac{2\Delta}{\kappa_{c}-\kappa_{ex}}\right]$.
Optical fields detuned from the cavity resonance are rotated by the cavity while optical fields on resonance with the cavity remains in phase inside and outside of the cavity.

\section{Probing the mechanics}

In order to determine the experimental conditions enabling optimal displacement sensing, we may first investigate the influence of detuning ($\Delta$) between the probe beam and the optical cavity resonance. In fact detuning has several detrimental effects on the measurement sensitivity. First of all Eq. (\ref{input-intra}) shows that increasing the detuning reduces the circulating power in the cavity and consequently weakens the optomechanical interaction strength $g$. Additionally the cavity induced phase rotations mix conjugate quadratures of the probe field into the output field what may hinder quantum-enhanced measurement with nonclassical states \cite{Hoff2013}. Detuning also affects the dynamics of the mechanical oscillator as the radiation pressure force that drives it depends on its position. This optomechanical feedback on the position of the mechanical oscillator modifies the stiffness and the damping rate of the mechanical resonator, a phenomenon that is known as dynamical backaction and may generate additional noise due to parametric instabilities \cite{Rokhsari2005}.

We now consider resonant probing of a cavity optomechanical system, i.e. $\Delta=0$, where all optical fields are in phase, thus $\tilde{X}^{\prime}=\tilde{X}$ and $\tilde{P}^{\prime}=\tilde{P}$. The amplitude and phase quadratures of the output field are derived from equations (\ref{OQLEf}) and (\ref{input-output}):
\begin{align}
\epsilon\tilde{X}_{out}&=\frac{2\sqrt{\eta(1-\eta)}}{1-2i\bar{\Omega}}\tilde{X}_{vac}+\frac{2\eta-1+2i\bar{\Omega}}{1-2i\bar{\Omega}}\tilde{X}_{d}, \label{XoutA}\\
\epsilon\tilde{P}_{out}&=\frac{2\sqrt{\eta(1-\eta)}}{1-2i\bar{\Omega}}\tilde{P}_{vac}+\frac{2\eta-1+2i\bar{\Omega}}{1-2i\bar{\Omega}}\tilde{P}_{d}-\frac{2g\sqrt{\eta/\kappa}}{1-2i\bar{\Omega}}\tilde{x},\label{XoutP}
\end{align}
where $\eta=\kappa_{ex}/\kappa$ is the escape efficiency of intracavity photons into the output field, and $\bar{\Omega}=\Omega/\kappa$ is the sideband frequency normalized to the cavity linewidth.

The motion of the mechanical oscillator is imprinted onto the phase quadrature of the output field. The power spectral density (PSD)
of phase quadrature fluctuations consists of the transduced mechanical oscillations added to the optical shot noise at sideband frequencies, and is given by
\begin{equation}\label{PSDout}
\tilde{S}_{out}^P=\left(1-\frac{4\eta(1-\eta)}{1+4\bar{\Omega}^2}\right)\tilde{S}_{d}^P+\frac{4\eta(1-\eta)}{1+4\bar{\Omega}^2}\tilde{S}_{vac}^P+\frac{4g^2\eta/\kappa}{1+4\bar{\Omega}^2}\tilde{S}_{x}.
\end{equation}
The output spectrum consists of three contributions. The first term in Eq. (\ref{PSDout}) comes from the probe field fluctuations, the second term arises due to cavity loss, and the third term represents the phase fluctuations induced by mechanical oscillations via the optomechanical interaction.
At sideband frequencies within the cavity linewidth ($\bar{\Omega}<1$), vacuum field fluctuations add to probe field fluctuations whereas at sideband frequencies far outside the cavity linewidth ($\bar{\Omega}\gg1$), shot noise is dominated by probe field fluctuations.

The optical shot noise of the output field adds an imprecision noise to the optomechanically transduced signal equivalent to mechanical oscillations with power spectral density:
\begin{equation}\label{imp_noise}
\tilde{S}_{x}^{imp}=\frac{\left(1+4\bar{\Omega}^2\right)\kappa^2}{64\eta^2 g_0^2\bar{s}_d^2}\left[\tilde{S}_{d}^P+\frac{4\eta(1-\eta)}{1+4\bar{\Omega}^2}\left(\tilde{S}_{vac}^P-\tilde{S}_{d}^P\right)\right],
\end{equation}
The imprecision noise is minimized when the optomechanical transduction is maximized. As the effective optomechanical interaction strength increases with the power circulating in the cavity, the smallest imprecision noise is achieved in the overcoupled regime ($\kappa_{ex}\gg\kappa_c$) within the cavity bandwidth ($\bar{\Omega} < 1$), and can be reduced further by increasing the probe power $P_d=\hbar \omega_d \bar{s}_d^2$.

In addition to imprecision noise due to the quantum noise of the probe field at all frequencies, random momentum kicks imparted onto the mechanical oscillator by fluctuations in the amplitude of the intracavity field drive mechanical oscillations, generating noise within the oscillator's linewidth called quantum backaction (QBA) noise. The QBA radiation pressure force impinging onto the mechanical oscillator and external forces, Eq. (\ref{MQLEf}), both drive the mechanical oscillator: 
\begin{equation}
\tilde{S}_x=\Omega_m^2\left|\chi_m\right|^2\left(\tilde{S}_{ex}+\tilde{S}_{qba}\right),
\end{equation}
where $\tilde{S}_{ex}$ and $\tilde{S}_{qba}$ are the spectra of the external and quantum backaction forces, respectively.
The power spectral density of QBA noise is obtained from Eq. (\ref{OQLEf}) at zero-detuning:
\begin{equation}
\tilde{S}_x^{qba}=\left|\chi_m\right|^2\frac{64\eta g_0^2\bar{s}_d^2\bar{\Omega}_m^2}{1+4\bar{\Omega}^2}\left[\tilde{S}_d^X+(1-\eta)\left(\tilde{S}_{vac}^X-\tilde{S}_d^X\right)\right].
\end{equation}
where $\bar{\Omega}_m=\Omega_m/\kappa$ is the oscillator's resonance frequency normalized to the optical cavity linewidth.
In contrast to the imprecision noise, QBA noise increases with the power circulating in the cavity due to increased radiation pressure backaction force driving the mechanical oscillator. Consequently QBA noise is minimized in the undercoupled regime  $\left(\kappa_{ex}\ll\kappa_c\right)$ outside of the cavity bandwidth $\left(\bar{\Omega} > 1\right)$. We may already foresee that there exists a trade off between imprecision noise and QBA noise for achieving  an optimal measurement sensitivity.

In the limit of zero ($\eta\rightarrow 0$) or infinite ($\eta\rightarrow 1$) cavity bandwidth, equivalent to the absence of a cavity, the optomechanical interaction is negligibly weak. Therefore the output phase quadrature consists only of phase noise from the probe field, which leads to infinite imprecision noise. Similarly the radiation pressure force from the probe field is negligibly weak, thus the QBA noise is negligibly small.

\begin{figure}[t]
  \includegraphics[width=\columnwidth]{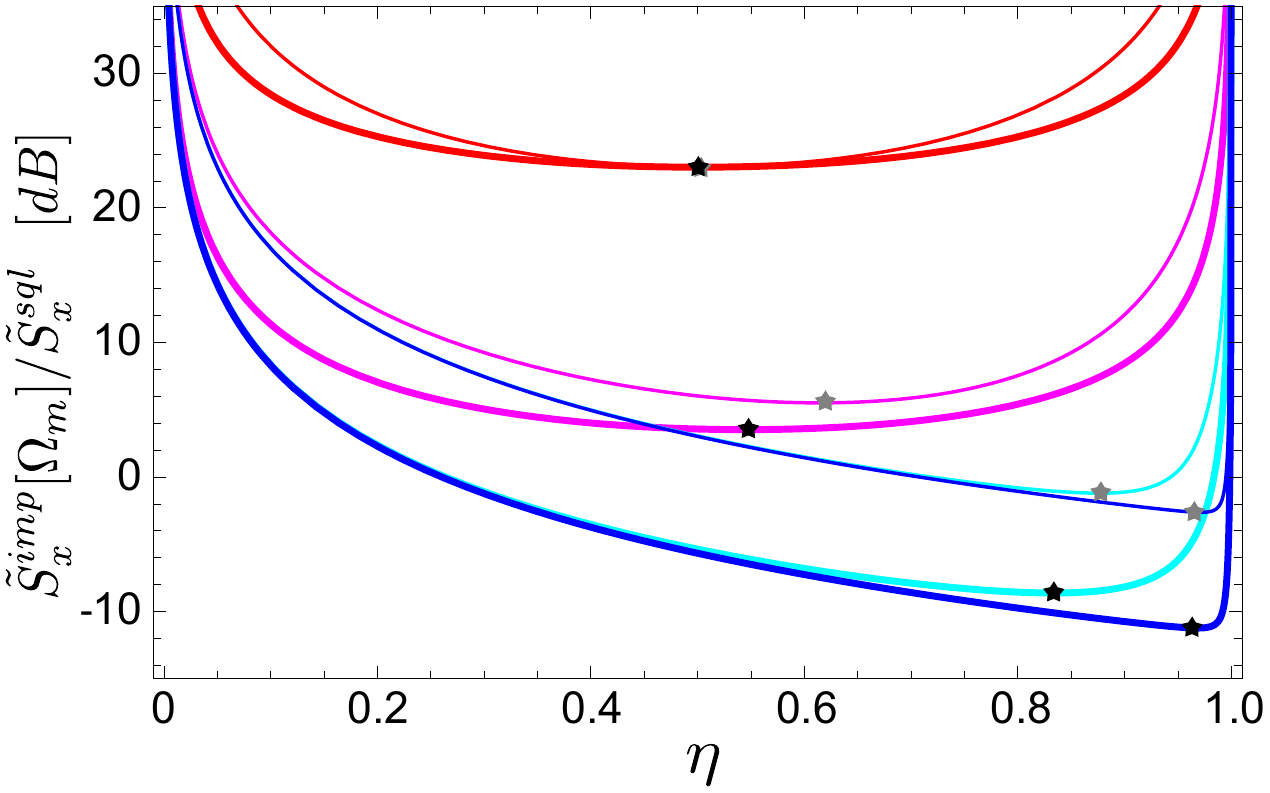}%
  \caption{\label{Simp_eta}
    Imprecision noise at the mechanical sideband for increasing sideband resolution $\Omega_m/\kappa_c=0.1$ (red), $1$ (magenta), $10$ (cyan) and $100$ (blue). Thin (thick) traces indicate imprecision noise for coherent (quantum-enhanced) probing with probe power $P_{min}$. Phase squeezed vacuum states with 8.7 dB reduced variance at the mechanical sidebands are used as quantum enhanced probing resources. Gray and Black stars  indicate the minimum of thin and thick traces, respectively.}
\end{figure}

Mechanical oscillations with angular frequency $\Omega_m$ scatter probe photons at $\omega_d$ into optical sidebands at $\omega_d\pm\Omega_m$. Therefore optical power measurements at these sidebands provides a measure of the amplitude of the mechanical oscillations. For resonant probing, information on the position of the oscillator is carried by the phase quadrature of the output field and can be retrieved, for example, by homodyne detection. We assume perfect detection efficiency for simplicity of the calculations. The total noise on the measurement of the mechanical oscillations' amplitude is given by the sum of imprecision noise and QBA noise at the mechanical sideband :
\begin{align}
\tilde{S}_x^{tot}[\Omega_m]
=&\tilde{S}_x^{imp}[\Omega_m]+\tilde{S}_x^{qba}[\Omega_m],\notag\\
=&\frac{1}{\sqrt{\eta}\Gamma_m\mathcal{P}}\left[\tilde{S}_d^P+\frac{4\eta(1-\eta)}{1+4\bar{\Omega}_m^2}\left(\tilde{S}_{vac}^P-\tilde{S}_{d}^P\right)\right]\notag\\
&+\frac{\mathcal{P}}{\sqrt{\eta}\Gamma_m}\left[\tilde{S}_d^X+\left(1-\eta\right)\left(\tilde{S}_{vac}^X-\tilde{S}_d^X\right)\right],\label{Stot}
\end{align}
where we have introduced the normalized power $\mathcal{P}$ : 
\begin{equation}
\mathcal{P}=\frac{P_{d}}{P_{opt}}=\frac{4\sqrt{\eta}}{1+4\bar{\Omega}_m^2}\frac{g^2}{\Gamma_m\kappa},
\end{equation}
which is the ratio of the probe power to an optimal power at which imprecision and QBA noises are balanced, thereby achieving a minimum in measurement sensitivity.

\begin{figure}[b]
  \includegraphics[width=\columnwidth]{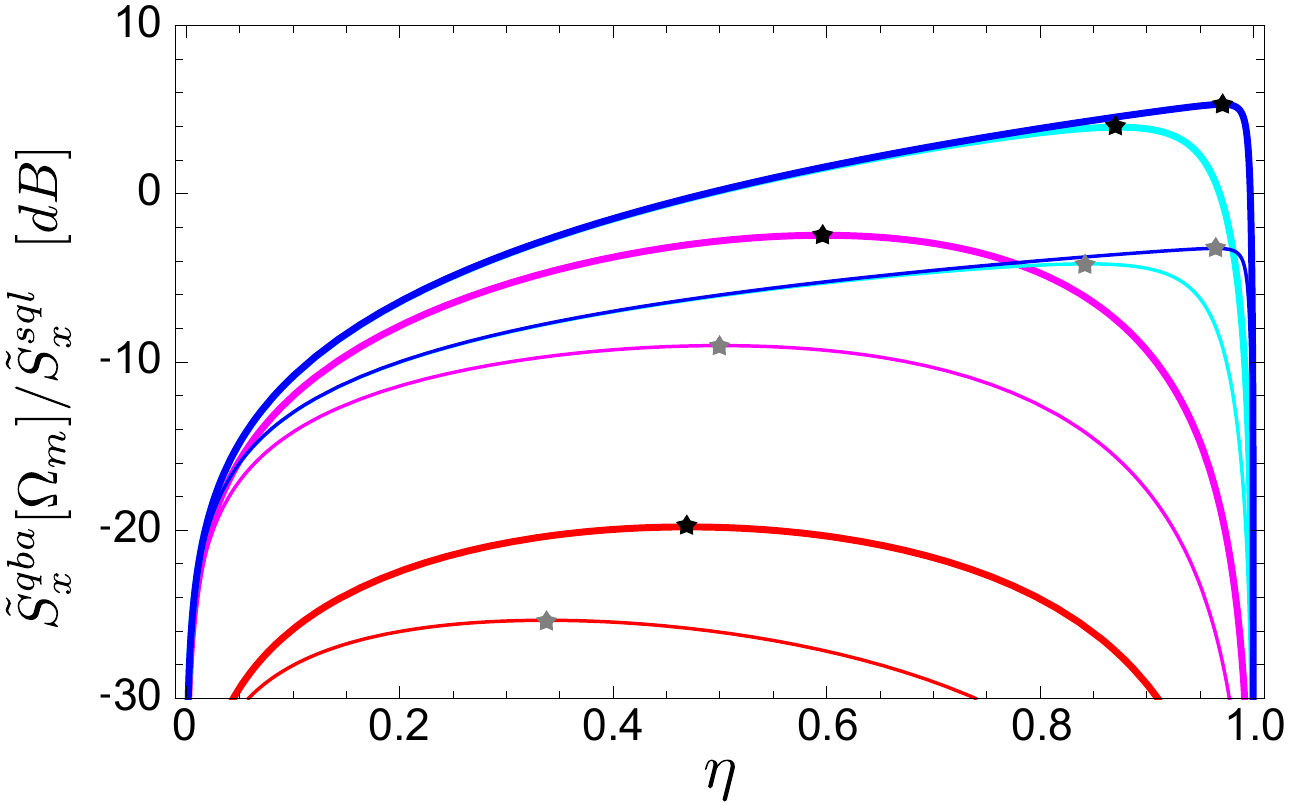}%
  \caption{\label{Sqba_eta}
    QBA noise at the mechanical sideband for increasing sideband resolution $\Omega_m/\kappa_c=0.1$ (red), $1$ (magenta), $10$ (cyan) and $100$ (blue). Thin (thick) traces indicate QBA noise for coherent (quantum-enhanced) probing with probe power $P_{min}$. Phase squeezed vacuum states with 8.7 dB reduced variance at mechanical sidebands are used as quantum enhanced probing resources. Gray and Black stars  indicate the maximum of thin and thick traces, respectively.}
\end{figure}

We consider two different probing resources to measure the oscillation amplitude of the mechanical resonator: coherent states and phase squeezed vacuum states. Coherent states obey the same correlation function as the vacuum state, i.e. $\tilde{S}_d^X=\tilde{S}_d^P=\tilde{S}_{vac}^X=\tilde{S}_{vac}^P=1$ \cite{Gardiner2004}.
Phase squeezed vacuum states result from quadrature entanglement between the sidebands of the probe field. Phase squeezed vacuum states are generated at the mechanical sideband frequencies, $\omega_d\pm\Omega_m$, within a bandwidth larger than the oscillator's linewidth, and obey the correlation function $\tilde{S}_d^P[\Omega_m]=1\left/\tilde{S}_d^X[\Omega_m]\right.=e^{-2r}$ \cite{Gardiner2004}, where $r\geq0$ is the degree of squeezing.
One requirement on implementing a squeezing-enhanced measurement scheme is to be able to generate significant phase squeezing at the mechanical sidebands.
Typically, mechanical oscillators used in quantum optomechanics experiments display resonance frequencies in the range of $1$-$100$ MHz. Several methods are also available to produce quadrature squeezed states in this frequency range, including optical parametric amplification, resonant second-harmonic generation and Kerr-nonlinearity in optical fibers.

\section{Limits of measurement sensitivity}

In order to determine the limits in measurement sensitivity of mechanical displacements we first consider the case where the cavity optomechanical system is driven by a coherent optical field. The total noise then reads $\tilde{S}_x^{coh}[\Omega_m]=\left(\mathcal{P}+1/\mathcal{P}\right)\left/\Gamma_m\sqrt{\eta}\right.$. The minimum in measurement sensitivity set by fundamental quantum noise is the standard quantum limit (SQL) and is reached for a normalized power equal to unity in the limit of high overcoupling, $\eta\rightarrow 1$ : $\tilde{S}_x^{coh}[\Omega_m]\rightarrow\tilde{S}_x^{sql}=2/\Gamma_m$.
Maximum sensitivity with a coherent probe field is achieved at a probe power $P^{sql}=\left[\eta^{-3/2}(1+4\bar{\Omega}_m^2)/4\bar{\Omega}_m^2\right]P_{min}$,
where $P_{min}\left/\hbar\omega_d\right. =\Gamma_m\Omega_m^2\left/16g_0^2\right.$ is the minimum probe power at which a measurement sensitivity at the SQL can be achieved. Reaching the SQL at a minimum probe power $P_{min}$ requires to be highly overcoupled ($\eta\rightarrow 1$) in the resolved sideband regime ($\bar{\Omega}_m\gg 1$).

Figures (\ref{Simp_eta}) and (\ref{Sqba_eta}) show imprecision noise and QBA noise at the mechanical sideband for a coherent and squeezed probe field with power $P_{min}$. In both figures the noise level is plotted against the cavity coupling parameter $\left(\eta=\kappa_{ex}\left/\kappa\right.\right)$ for various degrees of intrinsic sideband resolution $\left(\Omega_m\left/\kappa_c\right.\right)$ of the optomechanical system. A degree of squeezing $r=1$ is arbitrarily chosen, and yields a phase noise reduction on the probe field of $8.7$ dB, well within reach of current technology \cite{Eberle2010}. Both gray and black stars show that the minimum imprecision noise (maximum QBA noise) decreases (increases) as the resolution of the mechanical sideband increases, regardless of the probing method used. Probing mechanical oscillations at phase squeezed sidebands tends to reduce the imprecision noise while enhancing the QBA noise. The former trend is easily understood as imprecision noise arises from phase fluctuations of the probe beam while the later trend can be explained by anti-squeezed amplitude quadrature fluctuations driving mechanical oscillations.
It can be seen on Fig. (\ref{Simp_eta}) that quantum-enhanced probing yields limited improvement in measurement sensitivity in the bad-cavity limit $\left(\Omega_m\left/\kappa_c<1\right.\right)$ as intrinsic cavity loss damages squeezed states at sideband frequencies within the cavity linewidth.
When the cavity is critically coupled ($\eta=0.5$) and $\Omega_m \ll \kappa$, all drive fluctuations entering the cavity pass through and leave through loss ports. On the other hand, fluctuations that enter through loss ports all couple through the cavity to the output field. Therefore squeezing the drive field offers no benefit to imprecision noise reduction in this regime.
It is worthwhile here to draw an analogy to interferometry where the sensitivity is improved by injecting squeezed vacuum states in the dark port rather than the bright port \cite{Caves1981}. If vacuum fluctuations entering the cavity via $\kappa_c$ can be replaced by squeezed vacuum states then one can recover quantum enhanced sensitivity in the bad-cavity non-resolved sideband limit.

\begin{figure}[b]
  \includegraphics[width=\columnwidth]{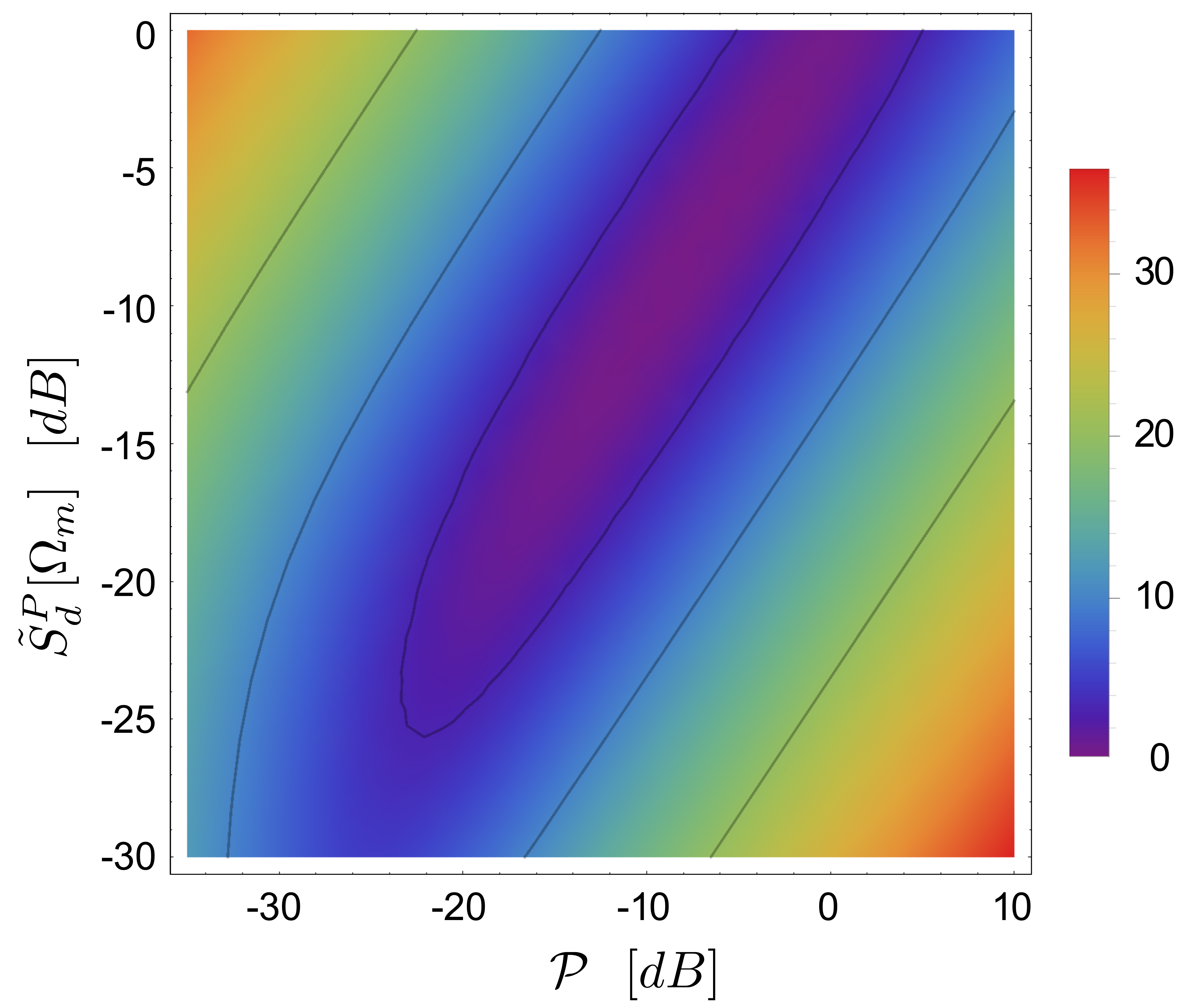}%
  \caption{\label{SQvsP}
    Quantum noise on mechanical displacement measurement in the overcoupled regime ($\eta=0.8$) at as a function of probe power and squeezing degree. The intrinsic sideband resolve parameter is $\Omega_m/\kappa_c =22$, corresponding to the system described in Ref. \cite{Verhagen2012}. Black lines indicate 3 dB, 10 dB and 20 dB of quantum noise above the SQL.}
\end{figure}

Optimum quantum-enhanced imprecision noise reduction is reached in the regime of negligible intracavity decay rate compared to both optical input coupling rate and mechanical resonance frequency, i.e. $\kappa_c \ll \kappa_{ex}, \Omega_m$, where the squeezed sidebands are far off resonance, and therefore little affected by cavity loss. Unfortunately this regime also corresponds to maximum QBA noise, although QBA noise is usually hidden by imprecision noise and Brownian motion, and has only been recently observed experimentally \cite{Purdy2013, Safavi2013}.

Fig. \ref{SQvsP} shows the total noise on measurements performed in the overcoupled regime ($\eta=0.8$) at various probe power and squeezing degrees. It is clear that strong squeezing of the phase quadrature fluctuations of the probe field gives access to measurement sensitivities close to the SQL at low powers. As expected from Eq. (\ref{Stot}), the minimum noise level is achieved for $\tilde{S}_d^P\left[\Omega_m]/\mathcal{P}\right.\approx1$. However, due to intrinsic cavity loss, i.e. $\eta<1$, the imprecision noise and QBA noise are unbalanced for high degrees of squeezing, what degrades the measurement sensitivity.

\section{Application to current systems}

\begin{figure*}[b]
  \includegraphics[width=\textwidth]{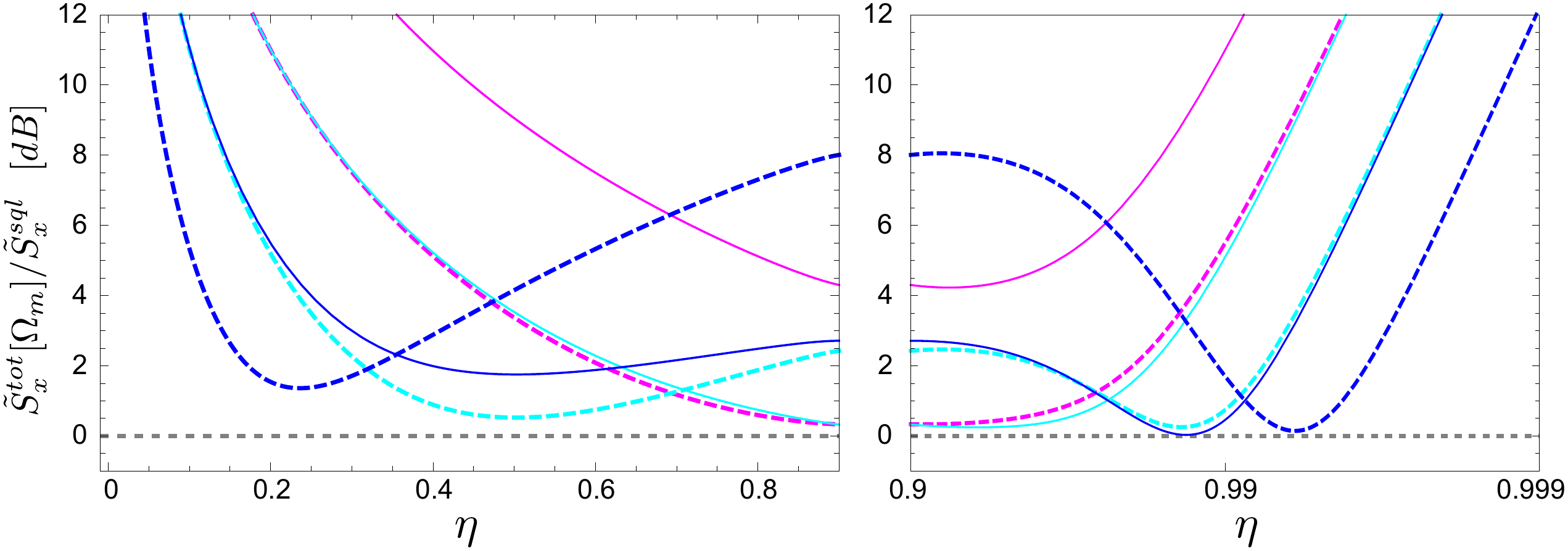}%
  \caption{\label{Stot_pow} 
  Quantum noise on mechanical displacement measurement with increasing probe power: -6 dB (magenta), 0 dB (cyan), and +6 dB (blue) over $P_{min}$. Plain traces correspond to coherent probing while dashed traces correspond to quantum-enhanced probing with phase noise reduction of 6 dB at sideband frequencies. The cavity optomechanical system is intrinsically in the resolved sideband regime with $\Omega_m/\kappa_c=22$ \cite{Verhagen2012}. The gray dashed trace represents the SQL.}
\end{figure*}

State-of-the-art cavity optomechanical systems that operate far into the resolved sideband regime $\left(\left.\Omega_m\right/\kappa\gtrsim 10\right)$ can be found among whispering-gallery-mode resonators \cite{Schliesser2008, Verhagen2012} and optomechanical crystals \cite{Eichenfield2009, Chan2011}. The later are not ideally suited for quantum-enhanced measurements as their high mechanical resonance frequencies ($\Omega_m>1$ GHz) lie outside of the bandwidth of state-of-the-art squeezing sources \cite{Ast2013}. The former display mechanical resonances at lower frequencies ($\Omega_m\sim100$ MHz) that can be probed with significant quantum noise reduction ($\sim 5$ dB @ $100$ MHz \cite{Ast2013, Mehmet2010}).

We consider the optomechanical system in Ref. \cite{Verhagen2012} which has a critically coupled sideband factor $\Omega_m/2\kappa_c =11$ for a mechanical resonance frequency of 78 MHz. Given optical and mechanical quality factors, $Q_c=5.5\times 10^7$ and $Q_m=9600$, and vacuum optomechanical coupling rate $g_0=2\pi\times 1.7 kHz$, we calculate the minimum probe power required to reach the SQL, $P_{min}=860$ nW. Fig. (\ref{Stot_pow}) shows the total noise on the measurement of mechanical oscillations induced by quantum fluctuations of the probe field, for probe powers : -6 dB (magenta), 0 dB (cyan), +6 dB (blue) over $P_{min}$. We compare quantum noise of displacement measurements for both coherent and quantum-enhanced probing. The squeezing source of Ref. \cite{Mehmet2010} can be used to provide phase squeezed vacuum states with 6 dB squeezing at 78 MHz sidebands. We consider the same degree of anti-squeezing for simplicity.
Clearly for impure squeezed states with increased anti-squeezing, the quantum backaction term would increase, degrading the optimal achievable sensitivity and precluding reaching the SQL.
As expected the measurement sensitivity increases with increasing probe power until the SQL is approached at $P_{min}$.
This is expected, since in the regime where imprecision noise dominates QBA noise, increasing the number of photons probing the mechanical resonator enhances the transduction of mechanical motion.
For measurements at probe powers above $P_{min}$, QBA noise is no longer negligible and may dominate imprecision noise above the SQL.
However, the QBA noise becomes less significant when the optical cavity is both strongly over- and under-coupled. Therefore the SQL can still be reached at higher optical coupling $\eta$. For a system intrinsically in the resolved sideband regime, probing the mechanics with a squeezed state improves the measurement sensitivity in a similar way as increasing probe power. For example, a measurement using a 10 dB phase-squeezed state achieves sensitivity equivalent to a 10 dB increase in probe power. This statement holds true when comparing measurement sensitivities at probe power lower than $P_{min}$ or in the strongly overcoupled regime. However, significant qualitative differences arise for measurements in the undercoupled regime at probe powers higher than $P_{min}$, where quantum-enhanced measurements perform better than an equivalent increase in probe power.

\section{Conclusion}

We have shown that optimum measurement sensitivities are reached in the highly overcoupled regime, $\eta\rightarrow 1$, for optomechanical systems with high intrinsic sideband factors, $\Omega_m\left/\kappa_c\right.\gg1$. High optomechanical coupling, $g_0$, and low mechanical damping, $\Gamma_m$, contribute to enhance the transduction of mechanical motion which has both the effect of reducing imprecision noise and increasing QBA noise, thereby lowering the probe power required to reach sensitivities at the SQL. At low probe power, keeping the QBA below shot noise, quantum-enhanced probing can significantly improve measurement sensitivities. On the other hand, quantum-enhanced probing yield little to no improvement in measurement sensitivities for systems with low intrinsic sideband factors, $\Omega_m\left/\kappa_c\right. <1$, in which optimal measurement sensitivities are reached at critical coupling. However quantum-enhanced measurements may find applications in systems where probe power is limited by technical or practical reasons. It may also provide a way to approach SQL in the undercoupled regime, where the finesse of the optical cavity is the highest.

\section*{Acknowledgments}

This research was funded by the Danish Council for Independent Research (Sapere Aude program) and the Lundbeck Foundation. W. P. Bowen acknowledges funding by the Australian Research Council Centre of Excellence CE110001013 and Discovery Project DP0987146.


\end{document}